\documentclass[12pt,a4j]{article}
\usepackage[dvips]{graphicx}
\usepackage{enumerate}
\usepackage{amsmath}

\begin{document}
\baselineskip=4.9mm
\centerline{\bf  Integrable multi-component generalization of a modified short pulse equation }\par

\bigskip
\centerline{Yoshimasa Matsuno\footnote{{\it E-mail address}: matsuno@yamaguchi-u.ac.jp}}\par

\centerline{\it Division of Applied Mathematical Science,} \par
\centerline{\it Graduate School of Sciences and Technology for Innovation} \par
\centerline{\it Yamaguchi University, Ube, Yamaguchi 755-8611} \par
\bigskip
\bigskip
We propose a multi-component generalization of the modified short pulse (SP) equation which was derived recently as
a reduction of Feng's two-component SP equation. Above all, we address the two-component system in depth.
We obtain the Lax pair,  an infinite nember of conservation laws and multisoliton solutions for the system, 
demonstrating its  integrability. Subsequently, we show that the  two-component system exhibits cusp solitons and breathers
for which the detailed analysis is performed. Specifically, we explore the interaction process of two cusp solitons and derive the formula for the phase shift.
While cusp solitons are singular solutions, smooth breather solutions are shown to exist, provided that the parameters characterizing the solutions satisfy certain condition.
Last, we discuss the relation between the proposed system and existing two-component SP equations. \par

 \par
\newpage
\leftline{\bf  I. INTRODUCTION} \par
\medskip
In a recent development of the theory of solitons, the short pulse (SP) equation has attracted considerable attention. 
It can be written in an appropriate dimensionless form as
$$u_{xt}=u+{1\over 6 }(u^3)_{xx}, \eqno(1.1)$$
where $u=u(x,t)$ represents the magnitude of the electric field and subscripts $x$ and $t$
appended to $u$ denote the partial differentiation.
The SP equation has been proposed as a model equation describing the propagation of ultrashort optical pulses in nonlinear media.$^{1}$
In the context of nonlinear optics, the cubic nonlinear Schr\"odinger (NLS)  equation has played a central role in studying 
the dynamics of optical solitons. While the NLS equation is applicable to the evolution of the slowly varying envelope, 
the SP equation works for short waves whose spectra are not localized around the carrier frequency.
A numerical analysis shows that as the pulse length shortens, the SP equation becomes a better approximation to
the solution of the Maxwell equation when compared with the prediction of the NLS equation.$^{2}$ We recall that the SP
equation has been derived for the first time in an attempt to constructing integrable partial differential equations (PDEs) 
associated with pseudospherical surfaces.$^{3, 4}$  The integrability aspects of the SP equation have been studied from various mathematical points of view.$^{5-7}$
Of particular interest is the existence of breather solutions whose characteristics are different from those of envelope soliton solutions of
the NLS equation.$^{8, 9}$  See  also a recent article which surveys the exact method of solution for the SP equation and the properties of
soliton and periodic solutions.$^{10}$
 \par
To take into account the effects of polarization or anisotropy,  the SP equation has been generalized to the  multi-component integrable systems.
Among them, Matsuno proposed the two-component system$^{11}$
$$u_{xt}=u+{1\over 2}(uvu_x)_x,\quad v_{xt}=v+{1\over 2}(uvv_x)_x, \eqno(1.2)$$
which is a special case of  the following coupled nonlinear PDEs for the $n$ variables $u_i=u_i(x, t), \ (i=1, 2, ..., n)$:
$$u_{i,xt}=u_i+{1\over 2}(Fu_{i,x})_x, \quad (i=1, 2, ..., n), \eqno(1.3a)$$
with
$$ \quad F={1\over 2}\sum_{1\leq j,k\leq n}c_{jk}u_ju_k. \eqno(1.3b)$$
Here, $c_{jk}$ are arbitrary constants with the symmetry $c_{jk}=c_{kj}  (j, k= 1,2, ..., n).$
Obviously, the identification $u_1=u$ and $u_2=v$ with $c_{11}=c_{22}=0, c_{12}=1$ in (1.3) yields (1.2).
\par
Another integrable generalization is due to Feng, which is given by$^{12}$
$$u_{xt}=u+{1\over 6 }(u^3)_{xx}+{1\over 2}v^2u_{xx},\quad v_{xt}=v+{1\over 6 }(v^3)_{xx}+{1\over 2}u^2v_{xx}. \eqno(1.4)$$
If we identify  $v$ with $u$, then the system (1.2) degenerates to the SP equation (1.1) while 
Feng's system (1.3) recasts to (1.1) upon putting $v=0$. \par
Quite recently, Sakovich considered the integrability of the nonlinear PDE$^{13}$
$$u_{xt}=u+au^2u_{xx}+buu_x^2, \eqno(1.5)$$
where $a$ and $b$ are arbitrary constants. Sakovich observed that the case $a/b=1/2$ corresponds to the SP equation (1.1) whereas
the case $a/b=1$ yields, after rescaling the variable $u$, a new  nonlinear PDE
$$u_{xt}=u+{1\over 2}u(u^2)_{xx}. \eqno(1.6)$$
Hereafter, we call Eq. (1.6) the modified SP equation. Note that Eq. (1.6) follows simply from Feng's system (1.4) by putting $v=\pm u$ and hence its
integrability will be assured.  \par
The main objective of the present paper is to generalize Eq. (1.6) to an integrable multi-component system.  Specifically, we propose the following coupled PDEs for the
$n$ variables $u_i=u_i(x,t)$:
$$u_{i,xt}=u_i+\left(Fu_{i,x}\right)_x-{1\over 2}\left(\sum_{1\leq j,k\leq n}c_{jk}u_{j,x}u_{k,x}\right)u_i,\quad (i=1, 2, ..., n), \eqno(1.7)$$
where $F$ is given by (1.3b).
For the special case of $n=2$,  we put $u_1=u$ and $u_2=v$ and $c_{11}=c_{22}=0, c_{12}=1$ and see that the system (1.7) reduces to  the two-component system
$$u_{xt}=u+v(uu_x)_x, \quad v_{xt}=v+u(vv_x)_x. \eqno(1.8)$$
Eq. (1.6) is obtained if one puts $v=u$ in (1.8) and hence one can regard the system (1.7) as a multi-component generalization of the modified SP equation. \par
The present paper is organized as follows. In Sec. II, we develop an exact  method for solving the modified SP equation. 
 Specifically, we employ a direct method (or the bilinear transformation method) combined with a hodograph transformation which worked well for the analysis of 
the SP equation.$^{9}$  We present the parametric representation of the multisoliton solutions, and show that they are closely related to the multisoliton solutions of
the sine-Gordon equation. The properties of solutions are briefly described. We also demonstrate that the modified SP equation transforms to the SP equation
through a hodograph transformation. In Sec. III, we generalize the modified SP equation to the multi-component system (1.7)  and provide its multisoliton solutions.
The basic strategy is to start from a multi-component version of the bilinear form associated with the modified SP equation and then transform it back to the system of equations for the
original variables through appropriate dependent variable transformations.
In Sec. IV, we analyze the two-component system (1.8) in detail.  We present its Lax pair, an infinite number of conservation laws and multisoliton 
solutions. We show that the system supports cusp solitons and breathers, and investigate their properties. 
Specifically, we address the interaction process of two cusp solitons, proving its solitonic feature.  Subsequently, we deal with the one-breather solution 
for which a condition for the existence of smooth breather is derived.
 Section V is devoted to concluding remarks. \par
\bigskip
\leftline{\bf II. MODIFIED SHORT PULSE EQUATION} \par
\medskip
In this section, we develop a systematic method for solving the modified SP equation (1.6).   While an
exact method has already been provided for Eq. (1.6),$^{13}$ we present an alternative approach which is applicable to  the multi-component system.
We also show that the SP equation is related to the modified SP equation through a hodograph transformation. \par
\medskip
\leftline{\bf A. Hodograph transformation} \par
\medskip
We introduce the hodograph transformation $(x ,t) \rightarrow (y, \tau)$ by
$$dy=rdx+ru^2dt, \quad d\tau=dt, \eqno(2.1a)$$
where $r(>0)$ is a function of $u$ to be determined later.
Then, the $x$ and $t$ derivatives transform as
$${\partial\over\partial x}=r{\partial\over\partial y},
\quad {\partial\over\partial t}={\partial\over\partial \tau}+ru^2{\partial\over\partial y}. \eqno(2.1b)$$
It turns out that  the variable $x=x(y, \tau)$ obeys a system of linear PDEs
$$x_y={1\over r},\quad x_\tau=-u^2. \eqno(2.2)$$
The solvability condition of this system, i.e., $x_{y\tau}=x_{y\tau}$ yields the evolution equation for $r$
$$r_\tau=2r^2uu_y. \eqno(2.3)$$
Applying the transformation (2.1) to the modified SP equation (1.6), one has
$$ru_{y\tau}+r_\tau u_y=u+r^2uu_y^2. \eqno(2.4)$$
If we multiply (2.4) by $u_y$ and eliminate the variable $u$ with the help of (2.3), we can put (2.4) into the form of  a linear ordinary differential equation for $u_y^2$
$${1\over 2}(u_y^2)_\tau+{r_\tau\over 2r}\,u_y^2={r_\tau\over 2r^3}. \eqno(2.5)$$
We impose the boundary conditions $u(\pm \infty, \tau)=0, r(\pm \infty, \tau)=1$. Then, Eq. (2.5) can be integrated with respect to $\tau$ to give the solution
$$u_y^2={1\over r}-{1\over r^2}. \eqno(2.6)$$
Using the relation $u_y=u_x/r$ which follows from (2.1b), we can determine the form of $r$ in terms of $u_x$
$$r=1+u_x^2. \eqno(2.7)$$
It follows from (2.3), (2.4) and (2.6) that
$$u_{y\tau}=\left({2\over r}-1\right)u=(2x_y-1)u. \eqno(2.8)$$
The equation (2.8) coupled with the system of equations (2.2) is the starting point in constructing multisoliton solutions. \par
\medskip
\leftline{\bf B. Parametric representation of multisoliton solutions} \par
\medskip
We construct the multisoliton solutions of the modified SP equation by means of the direct method.$^{14, 15}$ To this end, we introduce the following
dependent variable transformation for $u$ and the hodograph transformation for $x$
$$u={g\over f},  \eqno(2.9)$$
$$x=y+{h\over f}, \eqno(2.10)$$
where $f, g$ and $h$ are tau-functions which are the basic constituents of soliton solutions. The second equation in (2.2) is then transformed
to the bilinear equation
$$D_\tau h\cdot f=-g^2, \eqno(2.11)$$
whereas Eq. (2.8) reduces to
$${2gf_y\over f^3}(f_\tau+h)-{1\over f^2}(f_\tau g_y+f_yg_\tau+f_{y\tau}g+gh_y)+{1\over f}(g_{y\tau}-g)=0. \eqno(2.12)$$
Here, the bilinear operators $D_\tau$ and $D_y$ are defined by
$$D_\tau^mD_y^nf\cdot g=\left({\partial\over\partial \tau}-{\partial\over\partial \tau^\prime}\right)^m
\left({\partial\over\partial y}-{\partial\over\partial y^\prime}\right)^n
f(\tau, y)g(\tau^\prime, y^\prime)\Big|_{\tau^\prime=\tau,\, y^\prime=y},\quad (m, n = 0, 1, 2, ...).  \eqno(2.13)$$
We can decouple Eq. (2.12) into a set of equations
$$f_\tau+h=0, \eqno(2.14)$$
$$f_\tau g_y+f_yg_\tau+f_{y\tau}g+gh_y-f(g_{y\tau}-g)=0. \eqno(2.15)$$
Introducing $h$ from (2.14) into (2.15) and (2.11), we obtain the system of bilinear equations for $f$ and $g$:
$$D_yD_\tau f\cdot g=fg, \eqno(2.16)$$
$$D_\tau^2f\cdot f=2\,g^2. \eqno(2.17)$$
Then, the expression (2.10) with $h$ from (2.14) becomes
$$x=y-({\rm ln}\,f)_\tau, \eqno(2.18)$$
which, coupled with (2.9), gives a parametric representation of solutions.  The standard procedure in the context of the direct method$^{14, 15}$  can be applied to obtain soliton solutions
of the system of bilinear equations (2.16) and (2.17).  However, this problem will be discussed in Sec. IV. Below, we present an alternative approach.
\par
\medskip
\leftline{\bf C. Parametric solutions in terms of tau-functions of the sine-Gordon equation}\par
\medskip
It has been shown that the soliton solutions of the SP equation are expressed in terms of the tau-functions for the soliton solutions of the sine-Gordon equation.
We summarize the method developed in Ref. 9 keeping its application to the modified SP equation in mind.
First, solving (2.6) in $r$,  we find that $r$ has two roots
$$r={1\over 2u_y^2}\left(1\pm \sqrt{1-4u_y^2}\right). \eqno(2.19)$$
If we put
$$u_y={1\over 2}\sin\,\phi, \quad \phi=\phi(y, \tau),\eqno(2.20)$$
then (2.19) gives
$$r={1\over \cos^2{\phi\over 2}},\quad {1\over \sin^2{\phi\over 2}}. \eqno(2.21)$$
We choose the former solution, i.e., $r=1/\cos^2{\phi\over 2}$ and introduce this expression and (2.20) into (2.3) to express $u$ in terms of $\phi$
$$u={1\over 2}\phi_\tau. \eqno(2.22)$$
Substituting (2.22) into (2.20), we see that $\phi$ satisfies the sine-Gordon equation
$$\phi_{y\tau}=\sin\,\phi. \eqno(2.23)$$
\par
The $N$-soliton solution of Eq. (2.23) is expressed in terms of the tau-functions $f_{sG}$ and $\bar f_{sG}$ as
$$\phi=2{\rm i}\,{\rm ln}\,{\bar f_{sG}\over f_{sG}}, \eqno(2.24)$$
with
 $$ f_{sG}=\sum_{\mu=0,1}{\rm exp}\left[\sum_{j=1}^N\mu_j\left(\xi_j+{\pi\over 2}{\rm i}\right)
+\sum_{1\le j<k\le N}\mu_j\mu_k\gamma_{jk}\right], \eqno(2.25a)$$
  $$\bar f_{sG}=\sum_{\mu=0,1}{\rm exp}\left[\sum_{j=1}^N\mu_j\left(\xi_j-{\pi\over 2}{\rm i}\right)
+\sum_{1\le j<k\le N}\mu_j\mu_k\gamma_{jk}\right], \eqno(2.25b)$$
where
$$\xi_j=p_jy+{1\over p_j}\tau+\xi_{j0}, \quad (j=1, 2, ..., N),\eqno(2.26a)$$
$$e^{\gamma_{jk}}=\left({p_j-p_k\over p_j+p_k}\right)^2, \quad (j, k=1, 2, ..., N; j\not=k).
\eqno(2.26b)$$
Here, $p_j$ and $\xi_{j0}$ are arbitrary complex-valued  parameters satisfying the conditions $p_j\not=\pm p_k$
for $j\not= k$ and $N$ is an arbitrary positive integer. The notation $\sum_{\mu=0,1}$
implies the summation over all possible combinations of $\mu_1=0, 1, \mu_2=0, 1, ..., 
\mu_N=0, 1$. The tau-functions $f_{sG}$ and $\bar f_{sG}$ from (2.25) solve the bilinear equations
$$D_yD_\tau { f_{sG}}\cdot f_{sG}={1\over 2}( f_{sG}^2-{\bar f_{sG}}^2),
\quad D_yD_\tau {\bar f_{sG}}\cdot {\bar f_{sG}}={1\over 2}({\bar f_{sG}}^2- f_{sG}^2). \eqno(2.27)$$
It follows from (2.24) and (2.27) that
$$\cos\,\phi=1-2({\rm ln}\,\bar f_{sG}f_{sG})_{y\tau}. \eqno(2.28)$$
Inserting this expression into the relation 
$$x_y=1/r=\cos^2{\phi\over 2}={1\over 2}(1+\cos\,\phi), \eqno(2.29)$$
and then integrating (2.29) with respect to $y$, we obtain
$$x=y-({\rm ln}\,\bar f_{sG}f_{sG})_\tau+d, \eqno(2.30)$$
where $d$ is an integration constant depending generally on $\tau$.
If we substitute (2.22) with (2.24) and (2.30) into the second equation in (2.2), then we find
$$D_\tau^2\bar f_{sG}\cdot f_{sG}=d^\prime(\tau)\bar f_{sG}f_{sG}. \eqno(2.31)$$
We recall, however, that the tau-functions $f_{sG}$ and $\bar f_{sG}$ satisfy the bilinear equation
$$D_\tau^2{\bar f_{sG}}\cdot f_{sG}=0, \eqno(2.32)$$
and hence $d^\prime(\tau)=0$, showing that $d$ is independent of $\tau$. 
Due to the translational invariance of the modified SP equation, we can set this constant zero without loss of generality.
It follows from (2.22) and (2.24) that
$$u={\rm i}\left({\rm ln}\,{\bar f_{sG}\over f_{sG}}\right)_\tau.
 \eqno(2.33)$$
Consequently, (2.33) and (2.30) with the tau-functions (2.25) give a parametric representation for the $N$-soliton solution of the modified SP equation. \par
If we compare (2.9) and (2.18) with (2.33) and (2.30), respectively, we can infer that
$$f={\bar f_{sG}}f_{sG}, \quad g={\rm i}D_\tau {\bar f_{sG}}\cdot f_{sG}. \eqno(2.34)$$
We can verify these relations by showing that (2.34) indeed satisfy the bilinear equations (2.16) and (2.17).
Actually, inserting (2.27) into the identity
$$2D_yD_\tau[ab\cdot (D_\tau a\cdot b)]=D_\tau[a^2\cdot(D_yD_\tau b\cdot b)]+D_\tau[(D_yD_\tau a\cdot a)\cdot b^2], \eqno(2.35)$$
with $a={\bar f_{sG}}$ and $b=f_{sG}$, we obtain
$$2D_yD_\tau[{\bar f_{sG}} f_{sG}\cdot(D_\tau {\bar f_{sG}}\cdot f_{sG})]=D_\tau {\bar f_{sG}}^2\cdot f_{sG}^2=2{\bar f_{sG}}f_{sG}D_\tau {\bar f_{sG}}\cdot f_{sG}, \eqno(2.36)$$
which is just (2.16) with $f$ and $g$ from (2.34). To proceed, we use the identity
$$D_\tau^2 ab\cdot ab=2abD_\tau^2a\cdot b-2(D_\tau a\cdot b)^2, \eqno(2.37)$$
with $a={\bar f_{sG}}$ and $b=f_{sG}$. This gives
$$D_\tau^2{\bar f_{sG}}f_{sG}\cdot {\bar f_{sG}}f_{sG}=2{\bar f_{sG}}f_{sG}D_\tau^2{\bar f_{sG}}\cdot f_{sG}-2(D_\tau {\bar f_{sG}}\cdot f_{sG})^2. \eqno(2.38)$$
In view of (2.32), however, the first term on the right-hand side of (2.38)  vanishes, giving rise to (2.17). We remark that an alternative   proof of (2.34)
has been presented by employing a different bilinearization of the sine-Gordon equation from (2.27), i.e., $ D_\tau(D_\tau D_y-1){\bar f_{sG}}\cdot f_{sG}=0,\ D_\tau^2{\bar f_{sG}}\cdot f_{sG}=0.^{16}$
\par
\medskip
\leftline{\bf D. Transformation of the modified short pulse equation}\par
\medskip
Here, we show that the modified SP equation can be transformed to the SP equation via a hodograph transformation combined with a linear dependent variable transformation.
To this end, we introduce the transformations $u\rightarrow \bar u$ and $ (x, t)\rightarrow ({\bar x}, {\bar t})$
$$\bar u=2u, \quad \bar x=x-\int_{-\infty}^xu_x^2dx, \quad \bar t=t. \eqno(2.39)$$
Using the local conservation law $(u_x^2)_t=[u^2(1+u_x^2)]_x$ of the modified SP equation, we rewrite  the $\bar x$ and $\bar t$ derivatives in terms of $x$ and $t$ derivatives as
$${\partial\over \partial \bar x}={1\over 1-u_x^2}\,{\partial\over \partial  x}, \quad {\partial\over \partial \bar t}={\partial\over \partial  t}+{u^2(1+u_x^2)\over 1-u_x^2}\,{\partial\over \partial  x}. \eqno(2.40)$$
It is straightforward by applying (2.40) to the SP equation to show that
$$\bar u_{\bar x\bar t}-\bar u-{1\over 2}(\bar u^2\bar u_{\bar x})_{\bar x}={2\over 1-u_x^2}\left\{u_{xt}-u-{1\over 2}u(u^2)_{xx}\right\}. \eqno(2.41)$$
The relation (2.41) implies
 that if $u$ solves the modified SP equation, then $\bar u$ solves the SP equation, i.e.,
$$\bar u_{\bar x\bar t}=\bar u+{1\over 6}(\bar u^3)_{\bar x\bar 
x}, \eqno(2.42)$$
and vise versa. \par
We recall that the hodograph transformation $({\bar x}, {\bar t}) \rightarrow (\bar y, \bar \tau)$
$$d\bar y=\bar r\,d\bar x+{1\over 2}\bar r{\bar u}^2d\bar t, \quad d\bar\tau=d\bar t, \quad \bar r=\sqrt{1+{\bar u_{\bar x}}^2}, \eqno(2.43a)$$
or, equivalently in terms of the derivatives
$${\partial\over\partial \bar x}=\bar r{\partial\over\partial \bar y}, 
\quad {\partial\over\partial \bar t}={\partial\over\partial \bar \tau}+ {1\over 2}\bar r{\bar u}^2{\partial\over\partial \bar y}, \eqno(2.43b)$$
has been used for solving the SP equation.  It follows from  (2.40) and $\bar u=2u$  that $\bar u_{\bar x}=2(1-u_x^2)^{-1}u_x$. Substituting this relation into the definition of $\bar r$ from (2.43a) 
and referring to (2.7),  we obtain an important relation which connects $r$ with $\bar r$:
$$\bar r={r\over 2-r}. \eqno(2.44)$$
We use (2.1b), (2.40), (2.43b) and (2.44) to derive the relation
$$\bar r\,{\partial\over\partial \bar y}={r\over 1-u_x^2}\,{\partial\over\partial  y}={1+u_x^2\over 1-u_x^2}\,{\partial\over\partial  y}=\bar r\,{\partial\over\partial  y}. $$
Since $\bar r\not=0$, this gives $\partial/\partial \bar y=\partial/\partial y$.  Similarly, one has $\partial/\partial  \bar \tau=\partial/\partial  \tau$. 
It follows by applying these rules for the derivatives to Eq. (2.8) and then using (2.44) that
$$\bar u_{\bar y\bar\tau}={\bar u\over \bar r}. \eqno(2.45)$$
This equation coincides with the SP equation  (2.42) transformed by the hodograph transformation (2.43).
\par
\medskip
\leftline{\bf E. Soliton solutions }\par
\medskip
A few particular solutions of the modified SP equation such as soliton and breather have been found  in Ref. 13.  
 For completeness, we present the one-soliton solution in the framework of our approach and investigate its property. \par
 First, by solving the bilinear equations (2.16) and (2.17), we obtain the tau-functions for the one-soliton solution 
$$f=1+{\rm e}^{2\xi}, \quad g={2\over p}\,{\rm e}^{\xi},\quad \xi=py+{1\over p}\,\tau+\xi_0, \eqno(2.46)$$
where $p$ and $\xi_0$ are arbitrary real constants. The parametric representation of the one-soliton solution follows from (2.9) and (2.18). We write it in a convenient form
in the following analysis:
$$u={1\over p}\,{\rm sech}\,\xi, \eqno(2.47a)$$
$$ X\equiv x+ct-x_0={\xi\over p}-{1\over p}\,\tanh\,\xi, \eqno(2.47b)$$
where $c=1/p^2$ and $x_0=-(\xi_0+1)/p$. This represents a localized pulse with the amplitude $1/p$  moving to the left at the constant velocity $c$.
The $X$ derivative of $u$ can be computed from (2.47) to give
$$u_X={u_\xi\over X_\xi}=-{1\over \sinh\,\xi}. \eqno(2.48)$$
It turns out from (2.48) that $\lim_{X\rightarrow \pm 0}u_X=\mp\infty$. To examine the profile of the pulse at the crest $X=0$ (or $\xi=0$), we
expand $u$ and $X$ near the crest and obtain their leading-order asymptotics
$$u={1\over p}\left(1-{\xi^2\over 2}+O(\xi^4)\right), \quad
X={1\over p}\left({\xi^3\over 3}+O(\xi^5)\right). \eqno(2.49)$$
Eliminating the variable $\xi$ from (2.49) yields the profile of $u$ near the crest 
$$u={1\over p}\left(1-{1\over 2}(3pX)^{2/3}+O(X^{5/3})\right). \eqno(2.50)$$
Thus, the first derivative of $u$ with respect to $X$ does not exist at the crest, showing that $u$ takes the form of a cusp soliton,
as already observed in Ref. 13.   This  intriguing feature is in striking contrast to that of the SP equation for which its one-soliton solution is a loop soliton. 
\par
\bigskip
\leftline{\bf III. Multi-component generalization}\par
\medskip
\leftline{\bf A. Multi-component  bilinear system} \par
\medskip
To construct a multi-component analog of the modified SP equation, we start from the corresponding  bilinear equations and their parametric solutions.
Specifically, we generalize the parametric representation (2.9) and (2.18)  in the form
$$u_i={g_i\over f},\quad (i=1, 2, ..., n), \eqno(3.1a)$$
$$ x=y-({\rm ln}\,f)_\tau, \eqno(3.1b)$$
where the tau-functions $f$ and $g_i$ are assumed to satifiy the system of bilinear equations
$$D_yD_\tau g_i\cdot f=fg_i, \quad (i=1, 2, ..., n), \eqno(3.2)$$
$$D_\tau^2f\cdot f=\sum_{1\leq j, k\leq n}c_{jk}g_jg_k, \eqno(3.3)$$
which are the multi-component generalizations of (2.16) and (2.17), respectively.
Here, the coefficients $c_{jk}$ are the same as those  introduced in (1.3b). \par
\medskip
\leftline{\bf B. Multi-component modified short pulse equations} \par
\medskip
The next step is to rewrite (3.2) and (3.3) in terms of the variables $x$ and $t$. 
As will be demonstrated below, this can be achieved by means of
the hodograph transformation
$$dy=rdx+rFdt,\quad d\tau=dt, \eqno(3.4a)$$
or
$${\partial\over\partial y}={1\over r}{\partial\over\partial x},
\quad {\partial\over\partial \tau}={\partial\over\partial t}-F{\partial\over\partial x}. \eqno(3.4b)$$
where $F$ is given by (1.3b).  It turns out from (3.4) that the variable $x=x(y, \tau)$ satisfies the system of linear PDEs
$$x_y={1\over r}, \quad x_\tau=-F. \eqno(3.5)$$
The solvability condition of the system (3.5) yields the evolution equation for $r$
$$r_\tau=r^2F_y. \eqno(3.6)$$
\par
Now, we differentiate (3.1a) with respect to $y$ and $\tau$ and use (3.1a) and (3.2) to derive the relation
$$\left({g_i\over f}\right)_{y\tau}={D_yD_\tau g_i\cdot f\over f^2}-{(D_yD_\tau f\cdot f)g_i\over f^3}=u_i-{D_yD_\tau f\cdot f\over f^2}u_i,\quad (i=1, 2, ..., n). \eqno(3.7)$$
We substitute (3.1b) into the first equation in (3.5) to deduce it into the form
$${D_yD_\tau f\cdot f\over f^2}=2\left(1-{1\over r}\right). \eqno(3.8)$$
Upon introducing (3.1a) and (3.8) into (3.7), we find that
Eqs. (3.7) recast to
$$u_{i,y\tau}=\left({2\over r}-1\right)u_i, \quad (i=1, 2, ..., n). \eqno(3.9)$$
The above system of  equations is a multi-component analog of Eq. (2.8). 
We  rewrite (3.9) in terms of the original variables $x$ and $t$ by using (3.4b)
$$u_{i, xt}=(Fu_{i,x})_x+(2-r)u_i, \quad (i=1, 2, ..., n). \eqno(3.10)$$
\par
The last step is to determine the functional form of $r$. To this end, we use $F$ from (1.3b), (3.6) and (3.9) to derive the relation
$$\left({1\over r}\right)_\tau=-{r\over 2(2-r)}\sum_{1\leq j, k\leq n}c_{jk}(u_{j,y}u_{k,y})_\tau. \eqno(3.11)$$
Integrating (3.11) with respect to $\tau$ under the boundary conditions $u_{j,y}\rightarrow 0$ for  $j=
1, 2, ..., n$  and $ r\rightarrow 1$ as  $|y|\rightarrow \infty$, we obtain
$${1\over r}-{1\over r^2}={1\over 2}\sum_{1\leq j, k\leq n}c_{jk}u_{j,y}u_{k,y}. \eqno(3.12)$$
In view of the relations $u_{j,y}=u_{j,x}/r, u_{k,y}=u_{k,x}/r$ which stem from (3.4b), (3.12) becomes 
$$r=1+{1\over 2}\sum_{1\leq j, k\leq n}c_{jk}u_{j,x}u_{k,x}. \eqno(3.13)$$
Upon substituting (1.3b) and (3.13) into (3.10), we arrive at the following multi-component system which is a generalization of the modified SP equation:
$$u_{i,xt}=u_i+{1\over 2}\left[\left(\sum_{1\leq j, k\leq n}c_{jk}u_ju_k\right)u_{i,x}\right]_x
-{1\over 2}\left(\sum_{1\leq j, k\leq n}c_{jk}u_{j,x}u_{k,x}\right)u_i,\quad (i=1, 2, ..., n). \eqno(3.14)$$
\par
\medskip
\leftline{\bf C. Remarks} \par
\medskip
{\bf 1.}\ The  system of bilinear equations (3.2) and (3.3) coincides with 
that of the multi-component generalization of the SP equation proposed in Ref. 11 if one replaces the tau-functions $g_i$ by $2g_i\ (i=1, 2, ..., n)$.
 In accordance with this observation, the former system is found to exhibit the $N$-soliton solution for 
the following cases: $1)\ c_{j, k}\not=0\ (j\not=k), c_{jj}=0,\ (j, k=1, 2, ..., n),^{11} 2)\ c_{11}=c_{22}=1, c_{12}=c_{21}=0,^{11, 17, 18}
3)\ c_{jj}=1, c_{jk}=0\ (j\not=k; j, k=1, 2, 3, 4).^{18}$ 
\par
{\bf 2.}\ Solving (3.12) for $1/r$, we find
$${1\over r}={1\over 2}\left(1\pm\sqrt{1-2\sum\nolimits_{1\leq j, k\leq n}c_{jk}u_{j,y}u_{k,y}}\right). \eqno(3.15) $$
Introducing this expression into (3.9) while taking into account the boundary condition $r\rightarrow 1, |y|\rightarrow \infty$, we obtain a closed
system of PDEs for $u_i$
$$u_{i,y\tau}=\sqrt{1-2\sum\nolimits_{1\leq j, k\leq n}c_{jk}u_{j,y}u_{k,y}}\ u_i, \quad (i=1, 2, ..., n). \eqno(3.16)$$
\par
{\bf 3.}\  By means of the transformations $$\bar u_i=2u_i,\ (i=1, 2, ..., n), \quad \bar x=x-\int_{-\infty}^x(r-1)dx,\quad \bar t=t, \eqno(3.17)$$
we can derive the relation
$$\bar u_{i,\bar x\bar t}-\bar u_i-{1\over 2}(\bar F\bar u_{i,\bar x})_{\bar x}
={2\over 2-r}\left\{u_{i,xt}-u_i-(Fu_{i,x})_x-(1-r)u_i\right\},\quad (i=1, 2, ..., n),  \eqno(3.18)$$
which is a multi-component analog of (2.41), where
$$\bar F={1\over 2}\sum_{1\leq j, k\leq n}c_{jk}\bar u_j\bar u_k. \eqno(3.19)$$
Invoking Eqs. (3.14), we see that $\bar u_i$ satisfy the system of PDEs
$$\bar u_{i,\bar x\bar t}=\bar u_i+{1\over 2}(\bar F\bar u_{i,\bar x})_{\bar x}, \quad (i=1, 2, ..., n).  \eqno(3.20)$$
This system is a multi-component generalization of the SP equation proposed in Ref. 11. We can also derive the relation (2.44) between $r$ and $\bar r$,
where $r$ is given by (3.13) and  $\bar r$ by
$$ \bar r=\sqrt{1+{1\over 2}\sum\nolimits_{1\leq j, k\leq n}c_{jk}\bar u_{j,\bar x}\bar u_{k,\bar x}}. \eqno(3.21)$$
The hodograph transformation (2.43) with $\bar r$ from (3.21) and (3.4) recasts (3.9) to  the system of PDEs
$$\bar u_{i, \bar y\bar\tau}={\bar u_i\over \bar r}, \quad (i= 1, 2, ..., n), \eqno(3.22)$$
which is a multi-component generalization of Eq. (2.45). 
\par
{\bf 4.}\ If we put $c_{jj=1}\ (j=1, 2, ..., n), c_{jk}=0\ (j\not=k; j, k=1, 2, ..., n)$ in Eq. (3.14) and take the continuum limit $n\rightarrow \infty$, 
then we obtain a (2+1)-dimentional nonlocal PDE
$$u_{xt}=u+{1\over 2}\left(u_x\int^\infty_{-\infty}u^2dz\right)_x-{1\over 2}u\int^\infty_{-\infty}u_x^2dz,\quad u=u(x, z, t). \eqno(3.23)$$
The parametric representation for the solution of Eq. (3.23) can be expressed in the form 
$$u={g\over f}, \quad x=y-{f_\tau\over f}, \eqno(3.24)$$
where the tau-functions $f=f(y, \tau)$ and $g=g(y,z, \tau)$ satisfy the system of bilinear equations
$$D_yD_\tau f\cdot g=fg, \quad D_\tau^2f\cdot f=\int^\infty_{-\infty}g^2dz. \eqno(3.25)$$
The variables $y$ and $\tau$ in (3.24) and (3.25) are related to the original variables $x$ and $t$ by the hodograph transformation
$$dy=rdx+{1\over 2}\left(\int^\infty_{-\infty}u^2dz\right)r\,dt, \quad d\tau=dt, \eqno(3.26a)$$
where $r=r(x, t)$ is given by
$$r=1+{1\over 2}\int^\infty_{-\infty}u_x^2dz. \eqno(3.26b)$$
The integrable feature of Eq. (3.23) will be discussed elsewhere. \par
\bigskip
\leftline{\bf IV. Two-component system}\par
\medskip
In this section, we consider the two-component system (1.8) which is a special case of (3.14) with $n=2$ and $c_{11}=c_{22}=0, c_{12}=1$. 
We give the Lax pair, an infinite number of conservation laws and multisoliton solutions for the system, establishing its  integrability. \par
\medskip
\leftline{\bf A. Integrability} \par
\medskip
For the two-component system, Eqs. (3.5) and (3.9) become
$$x_{y\tau}=-(uv)_y,\quad u_{y\tau}=(2x_y-1)u, \quad v_{y\tau}=(2x_y-1)v, \eqno(4.1)$$
with the identifications $u_1=u, u_2=v$ and $F=uv$. In these expressions, $x_y=1/r$, where $r$
is determined by the relation
$${1\over r}-{1\over r^2}=u_yv_y. \eqno(4.2)$$
Note from (3.13) with $n=2$ that  $r$ is expressed in the $(x, t)$ coordinate system as
$$r=1+u_xv_x.\eqno(4.3)$$
By the reduction $u=v$,  Eqs. (4.1) reduce to Eqs. (2.2) and (2.8) whereas the expressions (4.2) and (4.3) reduce to (2.6) and (2.7), respectively.
We found that the system of equations (4.1) admits the following Lax pair
$${\Psi}_y=U{\Psi},\ {\Psi}_\tau=V{\Psi}, \eqno(4.4a)$$
with
$$U=\lambda\begin{pmatrix}2x_y-1 & 2u_y\\
                                    2v_y&-(2x_y-1)\end{pmatrix}, 
                                    \quad 
                                     V=\begin{pmatrix}{1\over 4\lambda}&-u\\ v&-{1\over 4\lambda}\end{pmatrix},
                                     \eqno(4.4b)$$
 where $\lambda$ is a spectral parameter.
Indeed, it follows from the compatibility condition $\Psi_{y\tau}=\Psi_{\tau y}$ that 
$$U_\tau-V_y+UV-VU=O, \eqno(4.5)$$
which yields Eqs. (4.1). \par
 The Lax pair associated with the system (1.8) is derived by rewriting (4.4) in terms of the
variables $x$ and $t$. This can be attained simply by applying the hodograph transformation (3.4) with $F=uv$ and $r$ from (4.3) to (4.4), giving
$${\Psi}_x=\tilde U{\Psi},\ {\Psi}_t=\tilde V{\Psi}, \eqno(4.6a)$$
with
$$\tilde U=\lambda\begin{pmatrix}2-r&2u_x\\
                                    2v_x&-(2-r)\end{pmatrix},
                                     \quad \tilde V=\begin{pmatrix}{1\over 4\lambda}+\lambda(2-r)uv&-u+2\lambda uvu_x\\
                                    v+2\lambda uvv_x&-{1\over4\lambda}-\lambda(2-r)uv\end{pmatrix}.    \eqno(4.6b)$$
It is easy to confirm that the compatibility condition of the Lax pair (4.6) yields the system of equations (1.8). \par
\medskip
\leftline{\bf B. Conservation laws} \par
\medskip
A striking feature of the completely integrable system is the existence of an infinite number of conservation laws.
Several practical methods are now available for deriving the conservation laws. Among them, we employ a procedure based on the Lax pair. See Ref. 19, for example. \par
Let $\Psi=(\psi_1, \psi_2)^T, \tilde U=(u_{ij})_{1\leq i, j\leq 2}, \tilde V=(v_{ij})_{1\leq i, j\leq 2}.$ Then, the Lax pair (4.6) is written in terms of the components as
$$\psi_{1,x}=u_{11}\psi_1+u_{12}\psi_2, \quad  \psi_{1,t}=v_{11}\psi_1+v_{12}\psi_2,
 \eqno(4.7a)$$
$$\psi_{2,x}=u_{21}\psi_1+u_{22}\psi_2, \quad \psi_{2,t}=v_{21}\psi_1+v_{22}\psi_2. \eqno(4.7b)$$
Introducing the varibles $\Gamma=\psi_2/\psi_1$ and $\bar\Gamma=\psi_1/\psi_2$, we can recast (4.7) to
$$({\rm ln}\,\psi_1)_x=u_{11}+u_{12}\Gamma, \quad ({\rm ln}\,\psi_1)_t=v_{11}+v_{12}\Gamma, \eqno(4.8a)$$
$$({\rm ln}\,\psi_2)_x=u_{22}+u_{21}\bar\Gamma, \quad ({\rm ln}\,\psi_2)_t=v_{22}+v_{21}\bar\Gamma. \eqno(4.8b)$$
It follows  from the compatibility conditions of (4.8) that
$$(u_{11}+u_{12}\Gamma)_t=(v_{11}+v_{12}\Gamma)_x, \eqno(4.9a)$$
$$(u_{22}+u_{21}\bar\Gamma)_t=(v_{22}+v_{21}\bar\Gamma)_x. \eqno(4.9b)$$
Integrating (4.9) with respect to $x$, we can see that the quantities $\int_{-\infty}^\infty(u_{11}+u_{12}\Gamma)dx,$\par 
\noindent $\int_{-\infty}^\infty(u_{22}+u_{21}\bar\Gamma)dx$
are conserved in time. We then use (4.7) to derive the relations
$$(u_{12}\Gamma)_x=u_{12}u_{21}+u_{12,x}\Gamma+(u_{22}-u_{11})u_{12}\Gamma-(u_{12}\Gamma)^2, \eqno(4.10a)$$
$$(u_{21}\bar\Gamma)_x=u_{12}u_{21}+u_{21,x}\bar\Gamma+(u_{11}-u_{22})u_{21}\bar\Gamma-(u_{21}\bar\Gamma)^2, \eqno(4.10b)$$
which reduce, after substituting the components of the matrix $\tilde U$, to
$$(u_{12}\Gamma)_x=4\lambda^2u_xv_x+{u_{xx}\over u_x}(u_{12}\Gamma)-2\lambda(2-r)(u_{12}\Gamma)-(u_{12}\Gamma)^2, \eqno(4.11a)$$
$$(u_{21}\bar\Gamma)_x=4\lambda^2u_xv_x+{v_{xx}\over v_x}(u_{21}\bar\Gamma)+2\lambda(2-r)(u_{21}\bar\Gamma)-(u_{21}\bar\Gamma)^2. \eqno(4.11b)$$
Note that (4.11a) transforms to (4.11b) by interchanging the variables $u$ and $v$ and replacing $\lambda$ by $-\lambda$.
This reflects the invariance of the system (1.8) under the interchange of the variables $u$ and $v$. 
Thus, we may use either (4.11a) or (4.11b) to obtain conservation laws. \par
Let us now derive the conservation laws.  We recall that $\int_{-\infty}^\infty(u_{11}+u_{12}\Gamma)_tdx=0.$
However, since $\int_{-\infty}^\infty u_{11,t}\,dx=0$, as confirmed easily,  $\int_{-\infty}^\infty u_{12}\Gamma\,dx$ is a conserved quantity.
We expand $u_{12}\Gamma$ in powers of $\lambda$ and substitute it into (4.11a). We find  two such expansions which will lead to both the local
and nonlocal conserved quantities.  We consider the two cases separately. \par
\medskip
\leftline{\bf 1. Local conservation laws}\par
\medskip
An expansion which yields the local conservation laws, i.e., integrals of the variables $u$ and $v$ and their $x$-derivatives,  is given by
$$u_{12}\Gamma=\sum_{n=0}^\infty \gamma_n\lambda^{1-n}. \eqno(4.12)$$
We impose the boundary conditions $u, v, u_x, v_x, ..., \gamma_n (n=0, 1, 2, ...) \rightarrow 0$ as $|x| \rightarrow \infty$ to assure 
the convergence of the integrals associated with the conservation laws.
Substituting (4.12) into (4.11a) and comparing the coefficients of $\lambda^{1-n}$ on both sides, we obtain the recursion relation that determines $\gamma_n$
$$\gamma_{n,x}={u_{xx}\over u_x}\,\gamma_n-2(2-r+\gamma_0)\gamma_{n+1}-\sum_{m=1}^n\gamma_{n-m+1}\gamma_m, \quad (n\geq 1), \eqno(4.13)$$
where $\gamma_0$ satisfies the quadratic equation
$$\gamma_0^2+2(2-r)\gamma_0-4u_xv_x=0, \eqno(4.14)$$
which stems from the coefficients of $\lambda^2$. It follows from the
 coefficients of order $\lambda$ that
$$\gamma_{0,x}={u_{xx}\over u_x}\,\gamma_0-2(2-r+\gamma_0)\gamma_1. \eqno(4.15)$$
With $r$ from (4.3), we find two solutions of Eq. (4.14), $\gamma_0=2u_xv_x$ and $\gamma_0=-2$. We substitute the former
solution into (4.15) to obtain
$$\gamma_1=-{u_xv_{xx}\over 1+u_xv_x}. \eqno(4.16)$$
Note that the solution $\gamma_0=-2$ is irrelevant since it does not satisfy the boundary condition.
 We put (4.13) in a form which is suitable for determinig $\gamma_n$  successively
$$\gamma_{n+1}={1\over 2(1+u_xv_x)}\left(-\gamma_{n,x}+{u_{xx}\over u_x}\,\gamma_n-\sum_{m=1}^n\gamma_{n-m+1}\gamma_m\right), \quad (n\geq 1). \eqno(4.17)$$
The $n$th conservation law which is denoted by  $I_n$ is  given by
$$I_n=\int_{-\infty}^\infty \gamma_ndx, \quad (n\geq 0). \eqno(4.18)$$
\par
The recursion relation (4.17) can be solved successively starting with the initial condition (4.16), the first two of which read
$$\gamma_2=-{1\over 2}{u_{xx}v_{xx}\over (1+u_xv_x)^3}+{1\over 2}\left({u_xv_{xx}\over (1+u_xv_x)^2}\right)_x, \eqno(4.19a)$$
$$\gamma_3={1\over 2(1+u_xv_x)}\left\{-\gamma_{2,x}+\left({u_{xx}\over u_x}+{2u_xv_{xx}\over 1+u_xv_x}\right)\gamma_2\right\}. \eqno(4.19b)$$
Although the computation of $\gamma_n$ becomes  very complicated as $n$ increases, it can be continued to generate explicit
expressions of $\gamma_n$. An inspection of  the structure of (4.17) reveals that the resulting conservation laws have local character.
Thus, up to overall constants, the first four conservation laws are found to be as follows:
$$I_0=\int_{-\infty}^\infty u_xv_xdx, \eqno(4.20a)$$
$$I_1=\int_{-\infty}^\infty{1\over r}(u_xv_{xx}-u_{xx}v_x)dx, \eqno(4.20b)$$
$$I_2=\int_{-\infty}^\infty {u_{xx}v_{xx}\over r^3}dx, \eqno(4.20c)$$
$$I_3=\int_{-\infty}^\infty\left[{1\over r^4}(u_{xx}v_{xxx}-u_{xxx}v_{xx})-{2\over r^5}u_{xx}v_{xx}(u_xv_{xx}-u_{xx}v_x)\right]dx.   \eqno(4.20d)$$
In deriving these formulas, we have used the fact that if $I_n$ is a conserved quantity, then the quantity $\tilde I_n$ which is obtained from $I_n$ by interchanging the
variables $u$ and $v$ is also conserved. Consequently, $I_n\pm \tilde I_n$ become the conservation laws as well. Note that $I_0$ and $I_2$ are symmetric and $I_1$ and $I_3$ 
are antisymmetric with respect to $u$ and $v$, respectively. \par
\medskip
\leftline{\bf 2. Nonlocal conservation laws}\par
\medskip
An expansion of $u_{12}\Gamma$ exists which produces nonlocal conservation laws.
To be more specific,  we expand $u_{12}\Gamma$ in powers of $\lambda$ as
$$u_{12}\Gamma=\sum_{n=1}^\infty \bar\gamma_n\lambda^n. \eqno(4.21)$$
Subsituting (4.21) into (4.11a) and comparing the coefficient of $\lambda^n$ on both sides, we obtain the recursion relation that determines $\bar\gamma_n$
$$\bar\gamma_1=u_x, \quad \bar\gamma_{n,x}=4u_xv_x\delta_{n,2}+{u_{xx}\over u_x}\,\bar\gamma_n-2(2-r)\bar\gamma_{n-1}-\sum_{m=1}^{n-1}\bar\gamma_{n-m}\bar\gamma_m, \quad (n\geq 2), \eqno(4.22)$$ 
where $\delta_{n,2}$ is Kronecker's delta. 
To solve the recursion relation for $\bar\gamma_n$, it is suitable to introduce the quantity $\hat\gamma_n$ by $\bar\gamma_n=u_x\hat \gamma_n$.
Thus,  (4.22) recasts to
$$\hat\gamma_1=1, \quad  \hat\gamma_{n,x}=4v_x\delta_{n,2}-2(2-r) \hat\gamma_{n-1}-u_x\sum_{m=1}^{n-1} \hat\gamma_{n-m} \hat\gamma_m,\quad (n\geq 2). \eqno(4.23)$$
This recursion relation can be solved successively with the initial condition $\hat\gamma_1=1$. Then, the $n$th conservation low  $J_n$ is given by the formula
$$J_n=\int_{-\infty}^\infty \bar\gamma_ndx=-\int_{-\infty}^\infty u \hat\gamma_{n,x}dx, \quad (n=1, 2, ...), \eqno(4.24)$$
where the last line follows by the integration by parts. \par
Before proceeding, we derive the useful relations which are a consequence of the system (1.8).  Specifically, we introduce the new variables $p$ and $q$ according to
$p=u(1-u_xv_x)$ and $q=v(1-u_xv_x)$ and put (1.8) into the form
$$u_{xt}=p+(uvu_x)_x\quad v_{xt}=q+(uvv_x)_x. \eqno(4.25)$$
It immediately follows from (4.25) and the boundary conditions $u, v, u_x, v_x \rightarrow 0, |x|\rightarrow \infty$ that
$$\int^\infty_{-\infty}p\,dx=\int_{-\infty}^\infty\,u(1-u_xv_x)dx=0, \quad \int^\infty_{-\infty}q\,dx=\int_{-\infty}^\infty\,v(1-u_xv_x)dx=0, \eqno(4.26)$$
and hence
$$\int_{-\infty}^\infty\,p_tdx=0,\quad \int_{-\infty}^\infty\,q_tdx=0. \eqno(4.27)$$
We use  Eqs. (4.25) integrated once with respect to $x$ to derive the evolution equations of $p$ and $q$. They read
$$p_t=(1-u_xv_x)\partial_x^{-1}p+2uvu_x-[u^2v(1+u_xv_x)]_x, \eqno(4.28a)$$
$$q_t=(1-u_xv_x)\partial_x^{-1}q+2uvv_x-[uv^2(1+u_xv_x)]_x, \eqno(4.28b)$$
where $\partial_x^{-1}=\int_{-\infty}^xdy$ is an integral operator.
If we substitute (4.28) into (4.27), we obtain the relations
$$\int_{-\infty}^\infty[(1-u_xv_x)\partial_x^{-1}p+2uvu_x]dx=0, \quad \int_{-\infty}^\infty[(1-u_xv_x)\partial_x^{-1}q+2uvv_x]dx=0. \eqno(4.29)$$
\par
Now, the recursion relation (4.23) yields the formulas
$$\hat\gamma_{2,x}=4v_x-2(2-r)\hat\gamma_1-u_x, \eqno(4.30a)$$
$$\hat\gamma_{3,x}=-2(2-r)\hat\gamma_2-2u_x\hat\gamma_2, \eqno(4.30b)$$
$$\hat\gamma_{4,x}=-2(2-r)\hat\gamma_3-2u_x\hat\gamma_3-u_x{\hat\gamma_2}^2,  \eqno(4.30c)$$
$$\hat\gamma_{5,x}=-2(2-r)\hat\gamma_4-2u_x\hat\gamma_4-2u_x\hat\gamma_2\hat\gamma_3.  \eqno(4.30d)$$
It is straightforward to obtain the conservation laws by substituting (4.30) into (4.24) and using the relations (4.26) and (4.29). 
 We omit the detail of the calculations and write only the final results. Up to overall constants, the first five conservation laws read
$$J_1=0, \eqno(4.31a)$$
$$ J_2=\int_{-\infty}^\infty( uv_x-u_xv)dx, \eqno(4.31b)$$
$$ J_3=\int_{-\infty}^\infty (1-u_xv_x)uvdx, \eqno(4.31c)$$
$$ J_4=\int_{-\infty}^\infty\Bigl[u^2vv_x-uu_xv^2+(1-u_xv_x)u\partial_x^{-1}\{(1-u_xv_x)v$$
$$  -(1-u_xv_x)v\partial_x^{-1}\{(1-u_xv_x)u\}\Bigr]dx, \eqno(4.31d)$$
$$J_5=\int_{-\infty}^\infty\Bigl[(1-u_xv_x)u^2v^2+2uvu_x\partial_x^{-1}\{(1-u_xv_x)v\}+2uvv_x\partial_x^{-1}\{(1-u_xv_x)u\}$$
$$+(1-u_xv_x)\partial_x^{-1}\{(1-u_xv_x)u\}\partial_x^{-1}\{(1-u_xv_x)v\}\Bigr]dx. \eqno(4.31e)$$
We can see that the conservation laws $J_n$ for  $n\geq 4$ become nonlocal in view of the presence of the integral operator $\partial_x^{-1}$. \par
Last, we conclude this section with a few comments.  First, we observe that under the reduction $u=v$, the conservation laws reduce to those
of the modified SP equation (1.6). Actually, the nontrivial conservation laws follows from (4.20) and (4.31):
$$I_0=\int_{-\infty}^\infty u_x^2dx, \eqno(4.32a)$$
$$I_2=\int_{-\infty}^\infty{u_{xx}^2\over (1+u_x^2)^3}dx, \eqno(4.32b)$$
$$J_3=\int_{-\infty}^\infty(1-u_x^2)u^2dx, \eqno(4.33a)$$
$$J_5=\int_{-\infty}^\infty\left[-{1\over 3}(1-u_x^2)u^4+(1-u_x^2)\left\{\partial_x^{-1}(1-u_x^2)u\right\}^2\right]dx. \eqno(4.33b)$$
\par
The second comment is concerned with the conservation laws of the two-component SP equation (1.2). As already shown in Sec. III, the transformations
$$\bar u=2u, \quad \bar v=2v, \quad \bar x=x-\int_{-\infty}^xu_xv_x\,dx, \quad \bar t=t, \eqno(4.34)$$
lead to (1.2) when applied to the two-component modified SP equation (1.8). This fact enables us to derive the conservation laws of the system (1.2) in
a simple manner. Indeed, using the relations $u_x=(1/2)(1-u_xv_x)\bar u_{\bar x}, v_x=(1/2)(1-u_xv_x)\bar v_{\bar x}$ and the two-component analog of (2.44),  the conservation laws $I_n$ and $J_n$ from
(4.20) and (4.31) reduce, up to overall constants, to $\bar I_n$ and $\bar J_n$, respectively, where
$$\bar I_0=\int_{-\infty}^\infty(\bar r-1)dx, \eqno(4.35a)$$
$$\bar I_1=\int_{-\infty}^\infty{1\over \bar r(\bar r+1)}(u_xv_{xx}-u_{xx}v_x)dx, \eqno(4.35b)$$
$$\bar I_2=\int_{-\infty}^\infty\left[{u_{xx}v_{xx}\over \bar r^3}-{\{(u_xv_x)_x\}^2\over 4\bar r^5}\right]dx, \eqno(4.35c)$$
$$\bar I_3=\int_{-\infty}^\infty{u_{xx}v_{xxx}-u_{xxx}v_{xx}\over \bar r^5}dx,  \eqno(4.35d)$$
$$\bar J_2=\int_{-\infty}^\infty(uv_x-u_xv)dx, \eqno(4.36a)$$
$$\bar J_3=\int_{-\infty}^\infty uv\,dx, \eqno(4.36b)$$
$$\bar J_4=\int_{-\infty}^\infty\left[u^2vv_x-uu_xv^2+4u\partial_x^{-1}v-4v\partial_x^{-1}u\right]dx, \eqno(4.36c)$$
$$\bar J_5=\int_{-\infty}^\infty\left[u^2v^2+2uv(u_x\partial_x^{-1}v+v_x\partial_x^{-1}u)+4(\partial_x^{-1}u)(\partial_x^{-1}v)\right]dx. \eqno(4.36d)$$
Here,  $\bar r=\sqrt{1+u_xv_x}$, and the overbar attached to the variables $u, v$ and $x$ has been  deleted for simplicity.  
Performing the reduction $u=v$ in the above expressions, we recover
 the conservation laws of the SP equation.$^{6, 7}$ \par
\medskip
\leftline{\bf C. Soliton solutions} \par
\medskip
The soliton solutions of the two-component modified SP equation are constructed by solving the system of bilinear equations
(3.2) and  (3.3) with $n=2$ and $ c_{11}=c_{22}=0, c_{12}=1$.  As already noticed, this system has the same form as that of the two-component SP equation (2.2).
For the latter system, the  tau-functions $f, g_1$ and $g_2$ for the $N$-soliton solution  have already been presented. See expressions (4.6) in Ref. 11.  
The only difference in the parametric representation of the solution lies in  the coordinate transformation.
Specifically, the expression corresponding to (2.18) takes the form  $x=y-2({\rm ln}\,f)_\tau$ for the two-component SP equation. 
We can construct multisoliton solutions in which each component contains arbitrary number of solitons.
Here, we restrict our consideration to the case where both $u$ and $v$ contain $N$ solitons.
We discuss the properties of the one- and two-soliton solutions, as well as the breather solution. The general $N$-soliton solution will be addressed shortly. \par
\medskip
\leftline{\bf 1. One-soliton solution}\par
\medskip
The tau-functions for the one-soliton solution are  given by
$$f=1+{a_1b_1p_1^2\over 4}z_1^2,\quad g_1=a_1z_1,\quad g_2=b_1z_1, \eqno(4.37a)$$
$$ z_1={\rm e}^{\xi_1}, \quad \xi_1=p_1y+{1\over p_1}\tau+\xi_{10}, \eqno(4.37b)$$
where $a_1, a_2$ and $p_1$  are real constants which are assumed to be positive here, and $\xi_{10}$ is  the phase constant.  
The parametric soliton solution  is calculated from (3.1) to give
$$u={1\over p_1}\sqrt{a_1\over b_1}\,{\rm sech}(\xi_1+\delta_1), \quad v={1\over p_1}\sqrt{b_1\over a_1}\,{\rm sech}(\xi_1+\delta_1), \eqno(4.38a)$$
$$x=y-{1\over p_1}\,{\tanh}(\xi_1+\delta_1),\quad \delta_1={\rm ln}\left({\sqrt{a_1b_1}p_1\over 2}\right). \eqno(4.38b)$$
\par
The  profile of $u$ is depicted in Fig. 1. It represents a cusp soliton with the amplitude ${1\over p_1}\sqrt{a_1\over b_1}$ 
and the velocity $c_1=1/ p_1^2$.  The property of $v$ is the same as that of $u$ except the amplitude and the velocity being given respectively 
by ${1\over p_1}\sqrt{b_1\over a_1}$ and $1/p_2^2$.
By comparing (2.47) and (4.38), we see that the cusp soliton has the same structure as that of the cusp soliton (2.47) of the modified SP equation. \par
\begin{center}
\includegraphics[width=8cm]{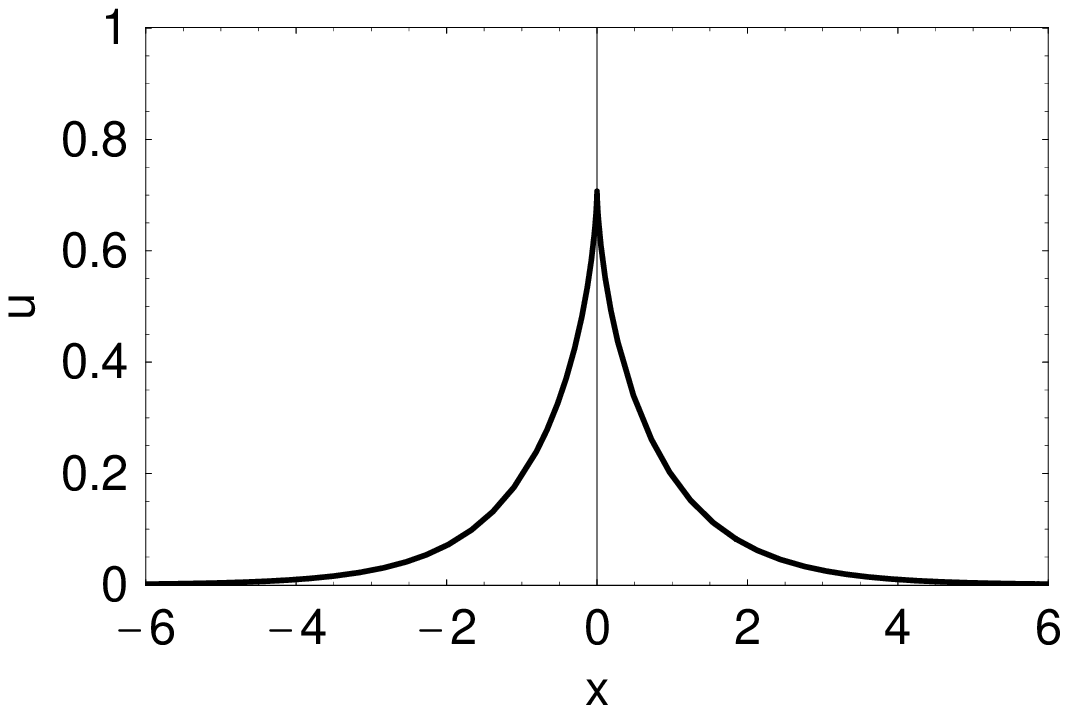}
\end{center}
{\bf FIG. 1.}\ The profile of a cusp soliton solution $u$  with the parameters $p_1=1.0, a_1=0.5$ and $b_1=1.0$. \par
\medskip
\leftline{\bf 2. Two-soliton solution}\par
\medskip
The tau-functions for the two-soliton solution read
$$f=1+{1\over 4}a_1b_1p_1^2z_1^2+(a_1b_2+a_2b_1){(p_1p_2)^2\over (p_1+p_2)^2}\,z_1z_2+{1\over 4}a_2b_2p_2^2z_2^2$$
$$+{1\over 16}a_1a_2b_1b_2{(p_1p_2)^2(p_1-p_2)^4\over (p_1+p_2)^4}\,(z_1z_2)^2, \eqno(4.39a)$$
$$g_1=a_1z_1+a_2z_2+{1\over 4}a_1a_2b_1{p_1^2(p_1-p_2)^2\over (p_1+p_2)^2}\,z_1^2z_2+{1\over 4}a_1a_2b_2{p_2^2(p_1-p_2)^2\over (p_1+p_2)^2}\,z_1z_2^2, \eqno(4.39b)$$
$$g_2=b_1z_1+b_2z_2+{1\over 4}a_1b_1b_2{p_1^2(p_1-p_2)^2\over (p_1+p_2)^2}\,z_1^2z_2+{1\over 4}a_2b_1b_2{p_2^2(p_1-p_2)^2\over (p_1+p_2)^2}\,z_1z_2^2. \eqno(4.39c)$$
\par
The asymptotic analysis of the two-soliton solution  can  be performed parallel to  that of the two-soliton solution of the two-component SP equation.  Hence, we summarize the results.
We are concerned only with $u$ since the corresponding  asymptotic formulas for $v$ follow by interchanging the parameters $a_i$ and $b_i$ $(i=1, 2)$.  \par
We decompose $u$ as $u=U_1+U_2$. Then, as $t\rightarrow -\infty$, $U_1$, $U_2$ and $x$ behave like
$$U_1 \sim {1\over p_1}\sqrt{a_1\over b_1}\, {\rm sech}(\xi_1+\delta_1^\prime), \eqno(4.40a)$$
$$x+c_1t-x_{10} \sim {\xi_1\over p_1}-{1\over p_1}\,{\rm tanh}(\xi_1+\delta_1^\prime)-{1\over p_1}-{2\over p_2},  \eqno(4.40b)$$
$$U_2\sim {1\over p_2}\sqrt{a_2\over b_2}\, {\rm sech}(\xi_2+\delta_2), \eqno(4.41a)$$
$$x+c_2t-x_{20} \sim {\xi_1\over p_2}-{1\over p_2}\,{\rm tanh}(\xi_2+\delta_2)-{1\over p_2},  \eqno(4.41b)$$
where
$$c_i={1\over p_i^2}, \quad \delta_i={\rm ln}\left({\sqrt{a_ib_i}\over 4}p_i\right), \quad \delta_i^\prime={\rm ln}\left[{\sqrt{a_ib_i}\over 2}p_i\left({p_1-p_2\over p_1+p_2}\right)^2\right],
    \ (i=1, 2). \eqno(4.42)$$
As $t\rightarrow +\infty$, the corresponding asymptotic forms  are given by
$$U_1 \sim {1\over p_1}\sqrt{a_1\over b_1}\, {\rm sech}(\xi_1+\delta_1), \eqno(4.43a)$$
$$x+c_1t-x_{10} \sim {\xi_1\over p_1}-{1\over p_1}\,{\rm tanh}(\xi_1+\delta_1)-{1\over p_1},  \eqno(4.43b)$$
$$U_2\sim {1\over p_2}\sqrt{a_2\over b_2}\, {\rm sech}(\xi_2+\delta_2^\prime), \eqno(4.44a)$$
$$x+c_2t-x_{20} \sim {\xi_1\over p_2}-{1\over p_2}\,{\rm tanh}(\xi_2+\delta_2^\prime)-{1\over p_2}-{2\over p_1}.  \eqno(4.44b)$$
We can see that the asymptotic state of the solution for large time is represented by a superposition  of two cusp solitons, each of which has the same form as that of
the cusp soliton solution given by (4.38).  The net effect of the interaction is the phase shifts caused by the collision of solitons. 
Let  $\Delta_1$ and $\Delta_2$ be the phase shifts of $U_1$ and $U_2$, respectively which are defined by
$$\Delta_i=x_{ic}(t\rightarrow  -\infty)-x_{ic}(t\rightarrow  +\infty), \quad (i=1, 2). \eqno(4.45)$$
where $x_{ic}$  is the center position of the $i$the soliton. It follows from (4.40)-(4.44) that
$$\Delta_1=-{1\over p_1}\,{\rm ln}\left({p_1-p_2\over p_1+p_2}\right)^2 -{2\over p_2}, \eqno(4.46a)$$
$$\Delta_2={1\over p_2}\,{\rm ln}\left({p_1-p_2\over p_1+p_2}\right)^2 +{2\over p_1}. \eqno(4.46b)$$
\par
 The asymptotic amplitudes of $U_1$ and $U_2$ for large time are given respectively
by  $A_1=(1/p_1)\sqrt{a_1/b_1}$ and $A_2=(1/p_2)\sqrt{a_2/b_2}$ and the velocities by $c_1=1/p_1^2$ and $c_2=1/p_2^2$, respectively.
Suppose that $A_1>A_2$ and  $c_1>c_2\ (p_1<p_2)$, implying that the large soliton travels faster than the small soliton.
  Figure 2 shows the interaction process  of two cusp solitons for the component $u$ with $A_1=2, A_2=1, c_1=4$ and $c_2=1$. 
Figure 3 plots $p_1\Delta_1$ and $p_1\Delta_2$ as a function of $s(=p_1/p_2)$.
We can see that the phase shift $\Delta_1$ of the large soliton is always positive whereas the small soliton exhibits a positive phase shift for $0<s<s_c$ and a negative phase
shift for $s_c<s<1$, where $s_c=0.648$ is a solution of the transcendental equation $\Delta_2=0$. In the present example, $\Delta_1=2.39$ and $\Delta_2=1.80$.
Recall that this peculiar feature of the phase shift has been observed in the interaction process of loop solitons of the SP and two-component SP equations.$^{9, 11}$ 
Another novel aspect of the solution is that the small soliton can travel faster than the large soliton if the condition $1<p_1/p_2<\sqrt{a_1b_2/a_2b_1}$ is satisfied 
so that $A_1>A_2$ and $c_1<c_2$.
\par

\begin{center}
\includegraphics[width=8cm]{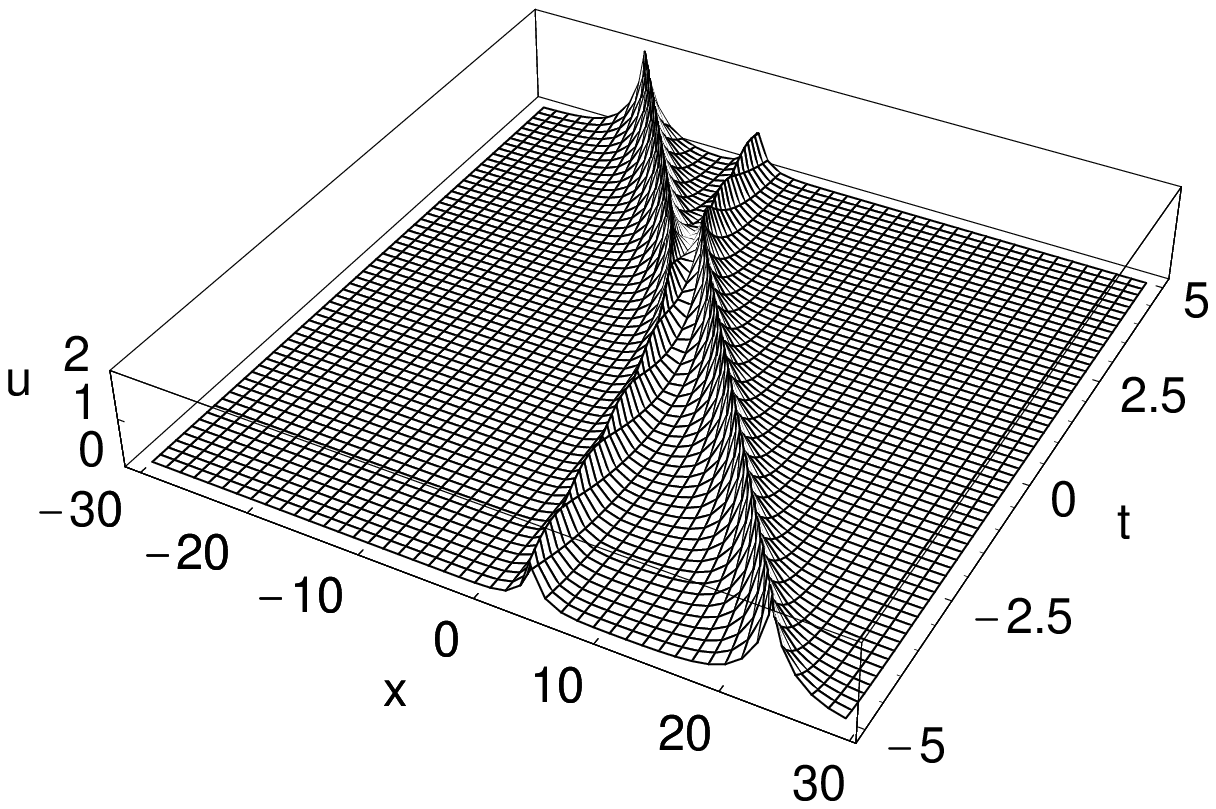}
\end{center}
{\bf FIG. 2.}\ The interaction process of the two-cusp soliton solution $u$ with the parameters $p_1=0.5, p_2=1.0, a_1=1.0, a_2=2.0, b_1=1.0, b_2=2.0$ and $\xi_{10}=\xi_{20}=0$. \par
\medskip
\begin{center}
\includegraphics[width=8cm]{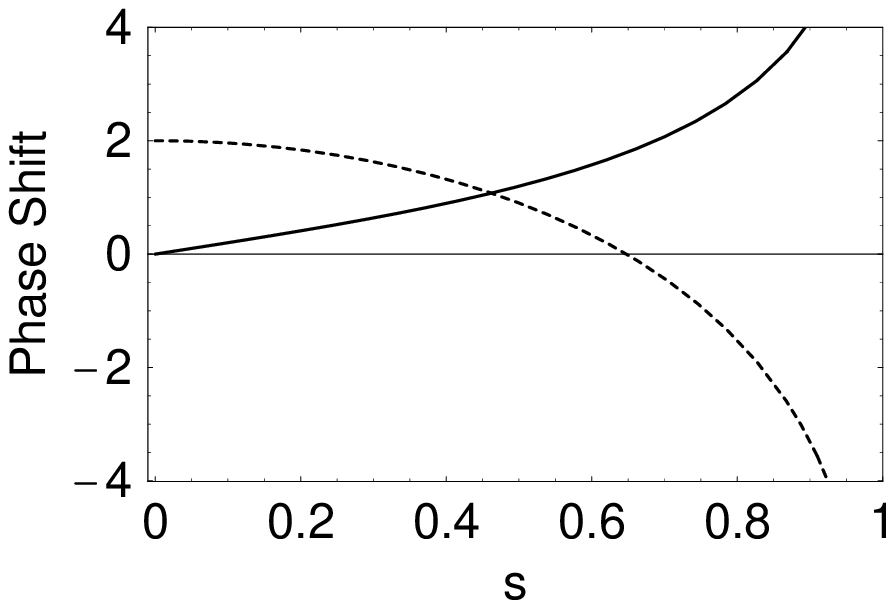}
\end{center}
{\bf FIG. 3.}\  The phase shifts $p_1\Delta_1$ and $p_1\Delta_2$ as a function of $s(=p_1/p_2)$. The solid (broken) line represents the phase shift of the large (small ) cusp soliton. \par
\medskip
\leftline{\bf 3. One-breather solution}\par
\medskip
The breather has a localized structure which oscillates with time and decays in space at infinity. Similar to the breather solution of the sine-Gordon 
equation which is the bound state of a kink and an antikink, the present two-component system supports breather solutions.
Of particular interest is the 
 one-breather solution which can be constructed from the two-soliton solution by means of a special parameterization.  Actually, we put 
$$p_1=a+{\rm i}b=p_2^*,\ \xi_{10}=\lambda+{\rm i}\mu=\xi_{20}^*,
\  a_1=\alpha{\rm e}^{{\rm i}\phi}=a_2^*, \ b_1=\beta{\rm e}^{{\rm i}\psi}=b_2^*, \eqno(4.47)$$
in (4.39) and obtain from (3.1) with $n=2$ the parametric representation of the solution
$$u={2ab\sqrt{\alpha\over \beta}\over\sqrt{a^2+b^2}}\,{\hat g_1\over\hat f},\quad v={2ab\sqrt{\beta\over \alpha}\over\sqrt{a^2+b^2}}\,{\hat g_2\over\hat f}, \eqno(4.48a)$$
$$x=y-{ab\over a^2+b^2}\,{b\,\sinh\,2\theta+a\,\sin\,2\chi\over \hat f}, \eqno(4.48b)$$
with
$$\hat f=b^2\cosh^2\theta+a^2\cos^2\chi-(a^2+b^2)\sin^2\delta, \eqno(4.49a)$$
$$\hat g_1=\sin(\chi_0-\delta)\,\sin\,\chi\,\cosh\,\theta-\cos(\chi_0-\delta)\,\cos\,\chi\,\sinh\,\theta, \eqno(4.49b)$$
$$\hat g_2=\sin(\chi_0+\delta)\,\sin\,\chi\,\cosh\,\theta-\cos(\chi_0+\delta)\,\cos\,\chi\,\sinh\,\theta, \eqno(4.49c)$$
$$\theta=a\left(y+{1\over a^2+b^2}\,\tau\right)+\lambda, \quad \chi=b\left(y-{1\over a^2+b^2}\,\tau\right)+\mu, \eqno(4.49d)$$
$$\tan\,\chi_0={b\over a},\quad \delta={1\over 2}(\phi-\psi). \eqno(4.49e)$$
Here, $a, b, \alpha$ and $\beta$ are positive constants, $\lambda, \mu, \phi$ and $\psi$ are real constants, the asterisk denotes complex conjugate, and
the appropriate shifts of the variables $x, y$ and $\tau$ have been performed to put the solution in a transparent form. 
Recall that the two-component modified SP equation (1.8) and its transformed form (4.1)  via the hodograph transformation (3.4) are
 invariant under the scale transformation $u\rightarrow u_0u, v\rightarrow v_0v, x\rightarrow b^{-1}x, y\rightarrow b^{-1}y,
t\rightarrow bt, \tau\rightarrow b\tau$ if the condition $u_0v_0b^2=1$ is inposed on the constants $u_0, v_0$ and $b$.
Applying this scale transformation to the solution (4.48) reveals that only the two parameters $a/b$ and $\delta$ characterize the solution.
The parameters $\lambda$ and $\mu$ in (4.49d) can be set to zero thanks to the translational invariance of (1.8).
\par
The breather solution presented here would exhibit singularities unless we impose certain condition on the parameters $a, b$ and $\delta$.
We shall address this issue. The smooth breather solution is produced if the inequalities $\hat f>0$ and $x_y>0$ hold for arbitrary values of $\theta$ and $\chi$.
The former condition is equivalent to requiring that the solution is finite in space and time, whereas the latter one is that the hodograph mapping (3.4) is one-to-one so that
the solution becomes a single-valued function of $x$ for arbitrary $t$.
The detailed analysis shows that the former inequality turns out to be $0<a/b<1/\sqrt{|\tan\,\delta|}$, whereas the latter one leads to the inequality
$$0<{a\over b}<\sqrt{1-|\sin\,\delta|\over 1+|\sin\,\delta|}, \quad -\pi\leq \delta\leq \pi. \eqno(4.50)$$
Since the permissible range of $a/b$ from (4.50) is included in the range from the former inequality, we impose (4.50) for the smooth breather solution to exist.
Figure 4 shows the time evolution of the one-breather solution $u$ where the parameters are given by $a=0.1, b=0.5, \alpha=\beta=1, \delta=\pi/4\, (\phi=\pi/2, \psi=0$).\par
\begin{center}
\includegraphics[width=8cm]{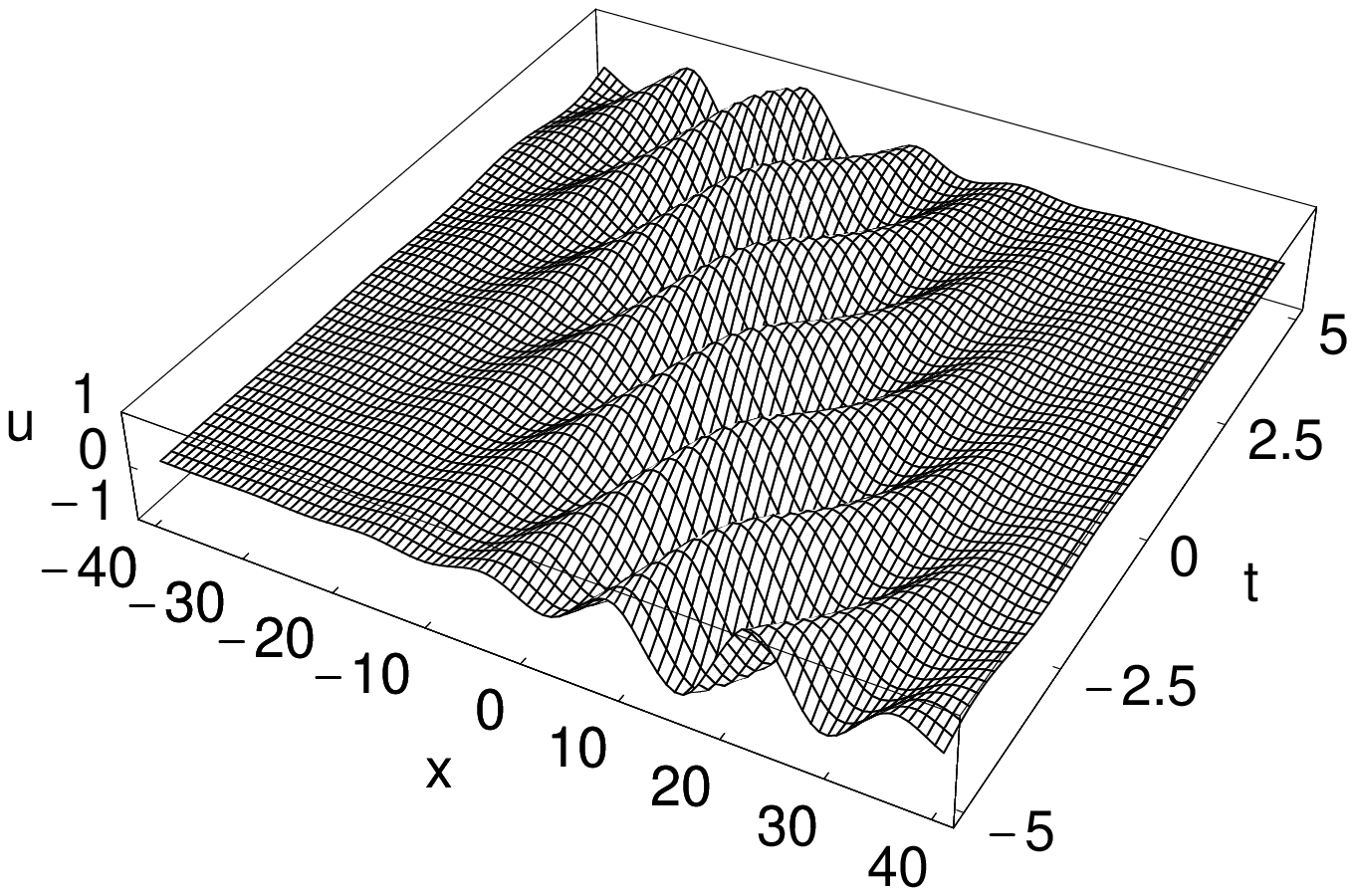}
\end{center}
\centerline{{\bf FIG. 4.}\ The time evolution of the one-breather solution $u$.} \par
\medskip
The properties of the solution depends critically on the parameter $\delta$. In particular, for $\delta=0$, the parametric solution (4.48) takes the form
$$u={2ab\sqrt{\alpha\over \beta}\over a^2+b^2}\,{b\, \sin\,\chi\,\cosh\,\theta-a\,\cos\,\chi\,\sinh\,\theta \over b^2\cosh^2\theta+a^2\cos^2\chi},\quad v={\beta\over\alpha}\,u, \eqno(4.51a)$$
$$x=y-{ab\over a^2+b^2}\,{b\,\sinh\,2\theta+a\,\sin\,2\chi\over b^2\cosh^2\theta+a^2\cos^2\chi}. \eqno(4.51b)$$
Under this special setting,  the inequality (4.50) becomes $0<a/b<1$, and the parameter $\chi_0$ from (4.49e) must satisfy the inequality $\tan\,\chi_0>1$.  In the above expression, we have assumed
$\pi/4<\chi_0<\pi/2$ so that $\cos\,\chi_0>0$. 
One can see from (4.51) that the solution describes the propagation of linearly polarized waves.
Note that if $\alpha=\beta$, then (4.51) reduces to the one-breather solution of the modified SP equation presented in Ref. 13.
In the limit of small amplitude $a/b\rightarrow 0$, (4.51) is approximated by the envelope soliton solution
$$u={2a\over b^2}\,{\sin\,\chi\over \cosh\,\theta}, \quad v={\beta\over\alpha}\,u, \eqno(4.52a)$$
$$x=y-{2a\over b^2}\tanh\,\theta. \eqno(4.52b)$$
\par
Another limiting value of $\delta=\pi/2$  deserves a  special attention. To derive a limiting form of the solution, let $\delta=\pi/2-\epsilon, (|\epsilon| \ll 1)$.
It turns out from (4.50) that the permissible values of $a/b$ lie in the range $0<a/b<|\epsilon|/2$.
Then, the parameter $\chi_0$ from (4.49) is approximated by $\chi_0\sim\pi/2 -a/b$. 
Inserting these values into (4.48), we obtain the leading order asymptotic of the parametric solution
$$u \sim {2a\over b^2}\sqrt{\alpha\over \beta}\,{\left(\epsilon-{a\over b}\right)\sin\,\chi\,\cosh\,\theta-\cos\,\chi\,\sinh\,\theta 
\over \sinh^2\theta-{a^2\over b^2}\sin^2\chi+\epsilon^2},  \eqno(4.53a)$$
$$v \sim {2a\over b^2}\sqrt{\beta\over \alpha}\,{\left(\epsilon+{a\over b}\right)\sin\,\chi\,\cosh\,\theta+\cos\,\chi\,\sinh\,\theta 
\over \sinh^2\theta-{a^2\over b^2}\sin^2\chi+\epsilon^2},  \eqno(4.53b)$$
$$x \sim y-{a\over b^2}\,{\sinh\,2\theta+{a\over b}\,\sin\,2\chi\over \sinh^2\theta-{a^2\over b^2}\sin^2\chi+\epsilon^2}. \eqno(4.53c)$$
If the parameter $a/b$ has the same order as $|\epsilon|$, then the amplitudes of $u$ and $v$ turn out to be of order 1. 
To be more specific, let $a/b=|\epsilon|\gamma$ with $0<\gamma<1/2$ and take the limit $\epsilon\rightarrow 0$. Then, the solution (4.53) tends to the limiting form
$$u = {2\over b}\sqrt{\alpha\over \beta}\,{\left(\gamma^{-1}\,{\rm sgn}\,\epsilon-1\right)\sin\,\hat\chi-\hat\theta\,\cos\,\hat\chi
\over {\hat\theta}^2-\sin^2\hat\chi+\gamma^{-2}},  \eqno(4.54a)$$
$$v = {2\over b}\sqrt{\beta\over \alpha}\,{\left(\gamma^{-1}\,{\rm sgn}\,\epsilon+1\right)\sin\,\hat\chi+\hat\theta\,\cos\,\hat\chi
\over {\hat\theta}^2-\sin^2\hat\chi+\gamma^{-2}},  \eqno(4.54b)$$
$$x=y-{1\over b}\,{2\hat\theta+\sin\,2\hat\chi\over{\hat\theta}^2-\sin^2\hat\chi+\gamma^{-2}},  \eqno(4.54c)$$
where $\hat\theta=b(y+\tau/b^2), \hat\chi=b(y-\tau/b^2)$ and we have put $\lambda=\mu=0$.  In the light of the scale invariance of the system (1.8) mentioned earlier,
the above solution is characterized  by a single parameter $\gamma$. 
\par
\medskip
\leftline{\bf 4. $N$-soliton solution}\par
\medskip
The $N$-soliton solution consists of a superposition of cusp solitons and breathers.  Let $n$ and $m$ be the number of cusp solitons and breathers, respectively.
Since the soliton parameters appear as  complex conjugate pairs for breathers,
this type of solutions is realized when  the condition $n+2m=N$ is satisfied. In particular, for pure breather solutions, we put $N=2m$ and impose the following conditions
to obtain the $m$-breather solution:
$$p_{2j-1}=p_{2j}^*=a_j+{\rm i}b_j, \quad a_j>0, \quad b_j>0, \quad (j=1,2, ..., m), \eqno(4.55a)$$
$$\xi_{2j-1}=\theta_j+{\rm i}\chi_j, \quad \xi_{2j}=\theta_j-{\rm i}\chi_j, \quad (j=1,2, ..., m), \eqno(4.55b)$$
$$a_{2j-1}=\alpha_j{\rm e}^{{\rm i}\phi_j}=a_{2j}^*, \ b_{2j-1}=\beta_j{\rm e}^{{\rm i}\psi_j}=b_{2j}^*, \quad (j=1,2, ..., m), \eqno(4.55c)$$  
where
$$\theta_j=a_j(y+c_j\tau)+\lambda_j, \quad \chi_j=b_j(y-c_j\tau)+\mu_j, \quad c_j={1\over a_j^2+b_j^2}, \quad (j=1,2, ..., m). \eqno(4.55d)$$  
The solution describes multiple collisions of $m$ smooth breathers provided that the condition similar to (4.50) is imposed on the parameters $a_j, b_j$
and $\delta_j(=(\phi_j-\psi_j)/2)\  (j=1,2, ..., m)$. 
This interesting issue is not discussed here and is left for a future study. \par
\medskip
\leftline{\bf D. Remarks} \par
\medskip
{\bf 1.}\ If we regard the variables $u$ and $v$ as complex-valued functions and impose the condition $v=u^*$, then the two-component system (1.8)  reduces to a single PDE for the
complex variable $w(\equiv u)$
$$w_{xt}=w+w^*(ww_x)_x, \quad w=w(x, t). \eqno(4.56)$$
This equation is a complex version of the modified SP equation. As will be understood, it becomes
 integrable.  
Actually, the Lax pair, conservation laws and soliton solutions of Eq. (4.56)  can be constructed simply  from the corresponding results
for Eq. (1.8) by the reduction $v=u^*$. In particular, the soliton solution is represented by the parametric representation 
$$w={g\over f},\quad x=y- ({\rm ln}\,f)_\tau, \eqno(4.57)$$
where the tau-functions $f$ and $g$ satisfy the system of bilinear equations
$$D_yD_{\tau}g\cdot f=fg, \quad D_\tau^2f\cdot f=2g^*g. \eqno(4.58)$$
Note in these expressions that $f$ should be a real-valued function. 
Equation (4.57) exhibits  breather solutions  as well as envelope solitons.  For example, the smooth envelope soliton takes the form
$$w={a\over a^2+b^2}\,{{\rm e}^{{\rm i}\chi}\over \cosh\,\theta}, \quad x=y-{a\over a^2+b^2}\,\tanh\,\theta, \eqno(4.59)$$
where $\theta$ and $\chi$ are given by (4.49d) and the condition $0<a/b<1$ is imposed to assure the smoothness
of the solution.  This solution describes the propagation of circularly polarized waves.
We remaind that the tau-functions for the $N$-soliton solution of Eqs. (4.58) have been presented in Ref. 11.
\par
{\bf 2.}\ The relation between Eq. (4.56) and the complex short pulse equation$^{11, 17, 18}$
$$q_{xt}=q+{1\over 2}(|q|^2q_x)_x, \quad q=q(x, t), \eqno(4.60)$$
is worth remarking.
One can see that  Eq. (4.60) is connected to Eq. (4.56) by the relation (3.18) 
through the transformations (3.17) with $n=2$, and the identification $w=u_1=u_2^*, q=\bar u_1=\bar u_2^*$. \par
{\bf 3.}\ As a model describing the unidirectional propagation of extremely short pulses in optical fibers, the following 
 equation has been proposed$^{20-23}$
$$E_{xt}=E+(|E|^2E)_{xx},\quad E=E(x, t), \eqno(4.61)$$
where $E$ represents a complex variable defined by $E=E_x+{\rm i}E_y$  with ${\bf E}=(E_x, E_y)$  being the electric field of the light wave.
While the only difference between Eq. (4.60) and Eq. (4.61) is the location of the $x$ derivative on the right-hand side, the structure
of solutions is widely different from each other. \par
Equation (4.61) admits an envelope soliton of the form
$$E={\sqrt{6}\over 9a}\,{{\rm e}^{{\rm i}\Phi}\over \cosh\left(ay+{t\over 9a}\right)}, \eqno(4.62a)$$
$$x=y-{2\over 3a}\,\tanh \left(ay+{t\over 9a}\right), \eqno(4.62b)$$
with
$$\Phi=2\sqrt{2}\left(ay-{t\over 9a}\right)+2\, \tan^{-1}\left[\sqrt{2}\,\tanh\left(ay+{t\over 9a}\right)\right], \eqno(4.62c)$$
where $a$ is an arbitrary nonzero constant. 
 In accordance with the invariance of Eq. (4.61) under the scale transformation 
$x\rightarrow a^{-1}x, t\rightarrow at, E\rightarrow a^{-1}E$,  this constant may be set to 1.
One can check by a direct substitution that (4.62) indeed satisfies Eq. (4.61).
When compared with an envelope soliton solution of Eq. (4.60) presented in Ref. 11 (see Eq. (4.31)), the solution (4.62) 
has a complex structure, particularly in the phase variable $\Phi$ which accounts for the occurrence of the strong phase modulation.
One interesting issue to be explored is whether Eq. (4.61) is  integrable or not. In this respect, we remark that it has passed the Painlev\'e
test, indicating an evidence of the integrability.$^{24}$
Nevertheless, the Lax pair, an infinite number of conservation laws and mulisoliton solutions have not been found for it 
which are common features to  integrable systems. \par
\bigskip
\newpage
\leftline{\bf V. CONCLUDING REMARKS}\par
\medskip
In this paper, we have shown that the modified SP equation admits an integrable multi-component generalization. In particular,
the  integrability of the two-component system has been established by constructing its Lax pair.
We have also demonstrated that the two-component modified SP equation is transformed to the two-component SP equation proposed by the author through a 
hodograph transformation. At the level of the bilinear equations,  both equations   are found to be
reduced to the same system of bilinear equations, the only difference being the coordinate transformation. However, this results in a new type of solutions.
We have addressed in detail  the properties of the   cusp solitons and  breather solutions
 for the  two-component modified SP equation in which we have derived a condition for the existence of the smooth one-breather solution.
 We can confirm the existence of smooth  multibreather solutions as well numerically by using the analytical solutions.  But, its rigorous proof still remains open.
 Another interesting issue is to generalize Feng's two-component system to the $n$-component system  with $n \geq 3$. 
 Although we have restricted our consideration to the mathematical aspects of the proposed system, 
the relevance of the system as a model capable of describing the dynamics of ultra-short pulses in optical fibers  is an important issue to be studied in a future work.$^{25}$
\par
\bigskip
\noindent{\bf ACKNOWLEDGEMENT} \par
\medskip
This work was supported partially by YAMAGUCHI UNIVERSITY FOUNDATION. \par

\newpage
\noindent {\bf REFERENCES}\par
\begin{enumerate}[1.]
\item T. Sch\"afer and C. E. Wayne, Propagation of ultra-short optical pulses in cubic nonlinear media, Physica D {\bf 196}, 90-105 (2004).
\item Y. Chung, C.K.R.T. Jones, T. Sch\"afer and C.E. Wayne, Ultra-short pulses in linear and nonlinear media, Nonlinearity {\bf 18}, 1351-1374 (2005).
\item  M. L. Rabelo, On equations which describe pseudospherical surfaces, Stud. Appl. Math. {\bf 81}, 221-248 (1989).
\item R. Beals, M. Rabelo and K. Tenenblat,  B\"acklund transformation and inverse scattering solutions for some pseudospherical surface equations, Stud. Appl. Math. {\bf 81}, 125-151 (1989).
\item A. Sakovich and S. Sakovich, The short pulse equation is integrable,  J. Phys. Soc. Jpn. {\bf 74}, 239-241  (2005).
\item J. C. Brunelli, The short pulse hierarchy, J. Math. Phys. {\bf 46}, 123507 (2005). 
\item J.C. Brunelli, The bi-Hamiltonian structure of the short pulse equation, Phys. Lett.  A {\bf 353}, 475-478 (2006).
\item A. Sakovich and S. Sakovich, Solitary wave solutions of the short pulse equation,  J. Phys. A {\bf 39}, L361-L367(2006).
\item Y. Matsuno, Multiloop soliton and multibreather solutions of the short pulse model equation,  J. Phys. Soc. Jpn. {\bf 76}, 084003 (2007).
\item Y. Matsuno, Soliton and periodic solutions of the short pulse model equation, {\it Handbook of Solitons: Research, Technology and Applications}, edited by S. P. Lang and H. Bedore (Nova, New York, 2009) pp. 541-585.
\item Y. Matsuno, A novel multi-component generalization of the short pulse equation and its multisoliton solutions, J. Math. Phys. {\bf 52}, 123702 (2011). 
\item B.-F. Feng, An Integrable coupled short pulse equation, J. Phys  A: Math. Theor. {\bf 45}, 085202 (2012).
\item S. Sakovich, Transformation  and integrability of a generalized short pulse equation, Commnu. Nonlinear Sci. Numer. Simulat. {\bf 39}, 21-28 (2016). 
\item Y. Matsuno, {\it Bilinear Transformation Method} (Academic Press, New York, 1984).
\item R. Hirota, {\it The Direct Method in Soliton Theory} (Cambridge University Press, 2004).
\item  R. Hirota and Y. Ohta, Hierarchy of coupled soliton equations. I, J. Phys. Soc. Jpn. {\bf 60}, 798-809 (1991).
\item A. Dimakis and F. M\"uller-Hoissen, Bidifferential calculas approach to AKNS hierarchies and their solutions, SIGMA {\bf 6}, 055 (2010).  
\item B.-F. Feng, Complex short pulse and coupled complex short pulse equations, Physica D {\bf 297}, 62-75 (2015).
\item M. Wadati, H. Sanuki and K. Konno, Relationships among inverse method, B\"acklund transformation and an infinite number of conservation laws,  Prog. Theor. Phys. {\bf 53}, 419-436 (1975).
\item D.V. Kartashov, A.V. Kim and S.A. Skobelov, Soliton structure of a wave field with an arbitrary number of oscillations in nonresonance media, JETP Lett. {\bf 78}, 276-280 (2003).
\item S.A. Skobelev, D.V. Kartashov and A.V. Kim, Few-optical-cycle solitons and pulse self-compression in a Kerr medium, Phys. Rev. Lett. {\bf 99}, 203902 (2007).
\item A.V. Kim, S.A. Skobelev, D. Anderson, T. Hansson and M. Lisak, Extreme nonlinear optics in a Kerr medium: Exact soliton solutions for a few cycles, Phys. Rev. A {\bf 77}, 043823 (2008).
\item Sh. Amiranashvili, A.G. Vladimirov and U. Bandelov, Solitary-wave solutions for few-cycle optical pulses, Phys. Rev. A {\bf 77}, 063821 (2008).
\item S. Sakovich, Integrability of the vector short pulse equation, J. Phys. Soc. Jpn. {\bf 77}, 123001 (2008).
\item  H. Leblond and D. Mihalache, Models of few optical cycle solitons beyond the slowly varying envelope approximation,  Phys. Rep. {\bf 523}, 61-126 (2013).

\end{enumerate}
\end{document}